\documentclass[12pt]{article}
\usepackage{amsmath,amsfonts,amsthm}
\usepackage{graphicx,psfrag,epsf,float}
\usepackage{enumerate}
\usepackage{natbib}
\usepackage{url} 
\usepackage{multirow}
\usepackage{longtable}
\usepackage{titling}
\usepackage{printlen}
\theoremstyle{definition}

\newtheorem{lemma}{Lemma}
\newtheorem{thm}{Theorem}

\newcommand{\E}      {\mbox{\tt E}}

\newcommand{\abold}    {\mbox{\bf a}}
\newcommand{\bbold}    {\mbox{\bf b}}

\newcommand{\ybold}    {\mbox{\bf y}}

\newcommand{\Gbold}    {\mbox{\bf G}}

\newcommand{\alphabold}   {\mbox{\boldmath${\alpha}$}}
\newcommand{\betabold}    {\mbox{\boldmath${\beta}$}}

\newcommand{\psibold}     {\mbox{\boldmath${\psi}$}}

\usepackage{xparse}
\usepackage{siunitx}
\usepackage{pgf}
\usepackage{mathtools}
 \usepackage{float}
 \usepackage{tikz}
 
\usepackage{natbib} 
\begin{document}
\def\spacingset#1{\renewcommand{\baselinestretch}%
{#1}\small\normalsize} \spacingset{1}
% \vspace{-2cm}
% \maketitle
% 13.70694cm
 % \convertto{in}{\textwidth} cm
% \printlength\textwidth
 \vspace{-3cm}
\title{ Recovering the Underlying Trajectory from Sparse and Irregular Longitudinal Data}
\author{Yunlong Nie, Yuping Yang and Jiguo Cao}
\date{Department of Statistics and Actuarial Science,\\ Simon Fraser University, Burnaby, BC, Canada}
 \vspace{-3cm}

\maketitle

\begin{abstract}
In this article, we consider the problem of recovering the underlying trajectory when the longitudinal data are sparsely and irregularly observed and noise-contaminated. Such data are popularly analyzed with functional principal component analysis via the Principal Analysis by Conditional Estimation (PACE) method. The PACE method may sometimes be numerically unstable because it involves the inverse of the covariance matrix. We propose a sparse orthonormal approximation (SOAP) method as an alternative. It estimates the optimal empirical basis functions in the best approximation framework rather than eigen-decomposing the covariance function. The SOAP method avoids estimating the mean and covariance function, which is challenging when the assembled time points with observations for all subjects are not sufficiently dense. The SOAP method avoids the inverse of the covariance matrix, hence the computation is more stable. It does not require the functional principal component scores to follow the Gaussian distribution. We show that the SOAP estimate for the optimal empirical basis function is asymptotically consistent.  The finite sample performance of the SOAP method is investigated in simulation studies in comparison with the PACE method. 
Our method is demonstrated by recovering the CD4 percentage curves from sparse and irregular data in the Multi-center AIDS Cohort Study. 
\end{abstract}
\newpage 
\spacingset{1.45} % DON'T change the spacing!
\section{Introduction}

Functional principal component analysis (FPCA) is a key dimension reduction tool in functional data analysis. FPCA explores major sources of variability in a sample of random curves by finding functional principal components (FPCs) that maximize the curve variation. Consequently, the top few FPCs explain most of the variability in the random curves. In addition, each random curve can be approximated by a linear combination of the top FPCs. Therefore, the infinite-dimensional curves are projected to a low-dimensional space defined by the top FPCs. This powerful dimensional reduction feature also contributes to the popularity of FPCA. 

The theoretical properties of FPCA have been carefully studied at length. For example,  \cite{dauxois1982asymptotic} first studied the asymptotic properties of PCA estimators for the infinite dimensional data from a linear operator perspective. Following this point of view, \cite{mas2002weak} and \cite{bosq2000linear} utilized functional analysis to study FPCA theoretically.  On the other hand, \cite{hall2007methodology,hall2006properties,yao2005functional} studied FPCA from the kernel perspective.  The smooth version of FPCA is carefully studied by \cite{rice1991estimating,pezzulli1993some,silverman1996smoothed,yao2005functional}. There are mainly three methods to achieve smoothness. The first method smooths the functional data in the first step and conducts the regular FPCA on the sample covariance function. The second method smooths the covariance function first and then eigen-decomposes the resulting smoothed covariance function to estimate the smoothed FPCs. The last method directly adds a roughness penalty in the optimization criterion for estimating FPCs.   

FPCA has been widely and successfully applied in many applications such as functional linear regression \citep{cardot1999functional}, classification and clustering of functional data (\cite{ramsay2006functional,muller2005functional,muller2005generalized,peng2008distance}). All these applications assume the functional data are densely and regularly observed. 

When the functional data are sparsely and irregularly observed, it is challenging to obtain a good estimate for FPCs and the corresponding FPC scores.  \cite{yao2005functional} proposed the Principal Analysis by Conditional Estimation (PACE) method to analyze the sparse functional data. The PACE method estimates the covariance function by the local polynomial regression method and then eigen-decomposes the estimated covariance function to obtain the eigenfunctions as the estimates of FPCs. The corresponding FPC score is estimated using conditional expectation, which requires that FPC scores follow a Gaussian distribution. The asymptotic properties are established in \cite{hall2006properties}. 

The PACE method is very successful. It is now popularly used to analyze sparse functional data. On the other hand, the PACE method also has two major assumptions, which may limit its applications. The first assumption of PACE is that the assembled time points with observations for all subjects are sufficiently dense. Otherwise, PACE cannot estimate the mean and covariance function by pooling data for all subjects together. The second assumption of PACE is that the FPC scores follow a Gaussian distribution. Otherwise, the conditional expectation formula is invalid. In addition, the PACE method involves the inverse of the estimated covariance matrix when estimating individual trajectories, which may be unstable. This problem will be demonstrated in our simulation studies. \cite{peng2009geometric} proposed a restricted maximum likelihood approach to estimate FPCs and apply a Newton-Raphson procedure on the Stiefel manifold to guarantee that the resulting FPCs satisfies the orthonormality constraints. They also used conditional expectation to obtain FPC scores in order to recover individual trajectories. Therefore, their method also involves the inverse of the estimated covariance matrix and requires the FPC scores to be Gaussian distributed. 

The main objective of this article is to recover the underlying trajectory given sparse and irregular longitudinal observations.  Note that this objective is different from exploring the major variation patterns of the functional data, which is the central goal for the conventional FPCA.  

We propose a new sparse orthonormal approximation (SOAP) method to recover the underlying trajectory.  This method directly estimates the optimal empirical basis functions and the corresponding coefficients in the best approximation framework.  The SOAP method has three main advantages. First, our method avoids the inverse of the covariance matrix, and the computation is stable and efficient. Second, it does not require that the scores follow the Gaussian distribution. Therefore, it can be applied in non-Gaussian cases. Lastly, our method does not need to estimate the mean and covariance function, which might be challenging when the assembled time points with observations for all subjects are not
sufficiently dense. The computing scripts for the simulation studies and the real data application are available in the supplementary file. 

The rest of the article is organized as follows. Section 2 introduces the best approximation framework for recovering the underlying trajectory given sparse and irregular longitudinal observations. Section 3 describes the SOAP method for estimating the optimal empirical basis functions and the corresponding coefficients. The asymptotic consistency results for the estimated functional empirical components(FECs) are provided in Section 4. Our proposed method is demonstrated in Section 5 by recovering the longitudinal CD4 percentage trajectories. In Section 6, we compare the finite sample performance of our method with the PACE method using simulation studies. Section 7 provides concluding remarks.

\section{Functional Empirical Component Analysis}

Consider $n$ independent realizations, $x_1(t),\ldots, x_n(t)$, of an $L^2$ stochastic process $X(t): t\in [0,T]$ at a sequence of random points on $[0, T]$ with measurement errors.  That is, the observed data $y_{ij}, i = 1\ldots, n, j = 1\ldots, n_i,$ is
$$y_{ij} = x_i(t_{ij})+\epsilon_{ij},$$
where $\{\epsilon_{ij}\}$ are independent and identically distributed random errors with  mean zero and variance $\sigma^2$. The number of measurements $n_i$ for each curve is random and small. The observed time points $t_{ij}$ can also be different for each curve.  Using the Karhunen-Lo\`eve expansion \citep{fukunaga1970representation},  each $x_i(t)$ can be expressed as
$$x_i(t) = \mu(t) + \sum_{k=1}^\infty\alpha_{ik}\phi_k(t),$$ 
where $\mu(t) = E(X(t))$ is the mean function, and $\phi_k(t), k = 1,2,\ldots,$ are the eigenfunctions of the covariance function $C(s, t) = E[(X(s) - \mu(s))(X(t) - \mu(t))], t, s\in [0, T]$. We call $\phi_k(t)$ the functional principal components (FPCs) and $\alpha_{ik}$ is the corresponding FPC score. The above estimation procedure is called the functional principal component analysis (FPCA).

A main advantage of FPCA is that $x_i(t)$ is projected to orthogonal basis functions, which allows us to approximate $x_i(t)$ using the first $K$ leading FPCs: 
$$x_i(t) \approx \mu(t) + \sum_{k=1}^K \alpha_{ik}\phi_k(t).$$
There are many other basis functions on which $x_i(t)$ can be projected. However, the eigenfunctions of the covariance functions have been  proved to be the optimal basis functions in the sense that they minimize the mean $L^2$ errors (see \cite{tran2008introduction}). Formally speaking, for any fixed $K \in\{1,2,\ldots\}$, the first $K$ leading FPCs minimize 
$$ \frac{1}{n}\bigg(\sum_{i=1}^n\int \bigg[x_i(t) -\mu(t)- \sum_{k=1}^K\langle x_i-\mu, \phi_k \rangle \phi_k(t)\bigg]^2dt\bigg)\,,$$
subject to $\langle\phi_k,\phi_l\rangle = \delta_{kl}$, where $\delta_{kl}$ is the Kronecker's delta. From the above criterion, we can see that the eigenfunctions $\phi_k(t), k=1,\ldots,K,$ are essentially the optimal empirical basis  functions to the centered stochastic process $X(t) - \mu(t)$. 

For the original stochastic process $X(t)$ without subtracting the mean function, the optimal empirical basis functions are the eigenfunctions of $K(s,t) = E[X(s)X(t)]$, as shown in Theorem 1 below.
% Due to this `optimal-basis-functions' properties, estimating the underlying $x_i(t)$ from its observations $\{y_{ij}, j=1,\ldots, n_i\}$, amounts to estimating the mean function $\mu(t)$, the $k$th eigenfunctions $\phi_k(t)$ and its corresponding scores $\alpha_{ki}$.  
Note that thought $K(s,t)$ is not a covariance function, it is a still Mercer kernel.  By Mercer's theorem, there exists an orthonormal basis $\psi_m(t)$  such that $K(s,t)$ has the following representation: 
$$K(s,t) = \sum_{m=1}^{\infty} \lambda_m \psi_m(s)\psi_m(t), $$
in which the eigenvalues $\lambda_1\geq\lambda_2\geq \ldots \geq 0$ and the eigenfunctions satisfy $\langle \psi_m , \psi_{\ell} \rangle =\delta_{m \ell}$. 
Correspondingly,  $x_i(t)$ can be represented as
$$x_i(t) = \sum_{m=1}^{\infty} \alpha_{im}\psi_m(t). $$
Now we will show that the empirical basis functions, $\psi_m(t), m=1, \ldots, M,$ optimal in the sense of minimizing the approximation error (\ref{th1}), are the eigenfunctions of the estimated $\widehat{K}(s,t) = \frac{1}{n}\sum_{i=1}^n[x_i(s)x_i(t)]$.
\begin{thm}\label{thm1}
For any given value of $M$, the optimal empirical basis functions $\psi_m(t), m=1, \ldots, M,$ which minimize 
\begin{align}
 \frac{1}{n} \sum_{i=1}^n\bigg(\int \bigg[x_i(t) - \sum_{m=1}^M \alpha_{im} \psi_m(t)\bigg]^2dt\bigg), \label{th1}
\end{align}
subject to $\langle \psi_m , \psi_{\ell} \rangle =\delta_{m \ell}$,  are the first $M$ eigenfunctions of  $\widehat{K}(s,t) = \frac{1}{n}\sum_{i=1}^n[x_i(s)x_i(t)]$ and $\alpha_{im} = \langle x_i,\psi_m \rangle$.
\end{thm}

The detailed proof for Theorem \ref{thm1} is given in the supplementary file. 
Theorem \ref{thm1} not only shows that those eigenfunctions of $\widehat{K}(s,t)$ are the optimal empirical basis functions to approximate the original functional data, but also provides an alternative way to estimate these optimal empirical basis functions in the best approximation framework other than eigen-decomposing the uncentered sample covariance function $ \widehat K(s,t)$. Note that estimating the sample covariance function may become challenging when the data are sparsely observed and the assembled time points with observations for all subjects are not sufficiently dense. 

Moreover, this best approximation framework also allows for estimating the coefficients to the optimal empirical basis functions without inverting the sample covariance matrix. Furthermore, Theorem \ref{thm1} shows that estimating the mean function $\mu(t)$ is not necessary if the goal is recovering or approximating the original trajectory. In practice, when the observed data are very sparsely observed and the assembled time points for all subjects are not sufficiently dense, it may be challenging to estimate the mean function $\mu(t)$. Alternatively,  we can simply estimate those optimal empirical basis functions and represent each trajectory using the estimated optimal empirical basis functions. 

In this article, the optimal empirical basis functions $\psi_m(t), m=1, 2, \ldots,$ are called the functional empirical components (FECs), and $\alpha_{im}$ is the corresponding FEC score. Note that when the mean function of the stochastic process $X(t)$, $\mu(t) = E(X(t)) = 0$, the functional empirical components are equivalent to the functional principal components. 

We propose the sparse orthonormal approximation (SOAP) method to estimate the first $M$ FECs $\psi_m(t), m=1, \ldots, M,$ by minimizing the observed loss function: 
\begin{align}
 \frac{1}{n} \sum_{i=1}^n\frac{1}{n_i}\sum_{j=1}^{n_i} \bigg[y_{ij} - \sum_{m=1}^M \alpha_{im} \psi_m(t_{ij})\bigg]^2, \label{sampleloss}
\end{align}
subject to $\langle \psi_m,\psi_{\ell} \rangle = \delta_{m \ell}$, where $m, \ell = 1,\ldots, M$. We solve the optimization problem \eqref{sampleloss} in a sequential manner. That is, we first obtain the first FEC. Then conditional on the estimated first FEC, we estimate the second FEC, and so on.  When estimating each FEC, we estimate the $m$-th component  $\psi_m$ and the corresponding FEC score $ \alphabold_m = (\alpha_{1m},\ldots,\alpha_{nm})^T$ in an iterative fashion.  We first estimate $\alphabold_m$ based on the given FEC $\psi_m(t)$ and the observations $y_{ij}, i=1,\ldots,n, j=1,\ldots,n_i$. Then,  given the estimated $\widehat{\alphabold}_m$, we obtain the corresponding FEC $\psi_m(t)$ by minimizing \eqref{sampleloss}. In each iteration, the loss function \eqref{sampleloss} is guaranteed to decrease. 

\section{Sparse Orthonormal Approximation Method}\label{EMFPC}
We first describe our sparse orthonormal approximation (SOAP) method to estimate the first FEC in Section \ref{k1}. Then our method is expanded to estimate the first $M$ FECs in Section \ref{k2}.

\subsection{Estimating the First FEC \label{k1}}
Based on \eqref{sampleloss}, the first FEC $\psi_1(t)$ is obtained by minimizing 
\begin{align}
\frac{1}{n} \sum_{i=1}^n\frac{1}{n_i}\sum_{j=1}^{n_i} \bigg[y_{ij}  -  \alpha_{i1}\psi_1(t_{ij})\bigg]^2, \label{loss1}
 \end{align}
 subject to $||\psi_1(t)||^2 = 1$. 
We first express $\psi_1(t)$ as a linear combination of basis functions: $\psi_1(t) = \betabold_1^T\bbold(t)$, where $\bbold(t) = (b_1(t),\ldots, b_L(t))^T$ is a vector of basis functions, and $\betabold_1 = (\beta_{11},\ldots,\beta_{1L})^T$ is the corresponding vector of coefficients. 
 We propose to minimize \eqref{loss1} in an iterative fashion. That is, for a given $\psi_1(t)$, we find the corresponding $\alpha_{i1}$ which minimizes \eqref{loss1}. Then given the value of $\alpha_{i1}$, we minimize \eqref{loss1} with respect to $\psi_1(t)$.  In every iteration step, the value of the lost function \eqref{loss1} decreases. The detailed algorithm is outlined as follows:
 \begin{enumerate}[Step I]
 \item Set the initial value of $\psi_1(t)$ as $\psi^{(0)}_1(t)$, which satisfies $||\psi_1^{(0)}||^2=1$;
 \item Given the current value of $\psi^{(\ell)}_1(t), j=0,1,2,\ldots$, we can obtain the value of $\alphabold^{(\ell)}_1 = (\alpha_{11},\ldots,\alpha_{n1})^T$ by minimizing 
 $$\frac{1}{n} \sum_{i=1}^n\frac{1}{n_i}\sum_{j=1}^{n_i} \bigg[y_{ij}  -  \alpha_{i1}\psi^{(\ell)}_1(t_{ij})\bigg]^2.$$
 In fact, this is simply a least squares problem. The $i$th element of $\alphabold^{(\ell)}_1$ can be expressed as 
 $$\alpha_{i1} =(\psibold_{1i}^T\psibold_{1i})^{-1}\psibold_{1i}^T \ybold_i. $$
 where $\psibold_{1i} = (\psi_1(t_{i1}),\ldots, \psi_1(t_{in_i}))^T$ is a $n_i\times 1$ vector and $\ybold_i = (y_{i1},\ldots, y_{in_i})^T$. 
  \item Given the current value of $\alphabold^{(\ell)}_1$, we update $\psi^{(\ell)}_1(t)$ to  $\psi^{(\ell+1)}_1(t)$ by minimizing
 $$ \frac{1}{n} \sum_{i=1}^n\frac{1}{n_i}\sum_{j=1}^{n_i} [y_{ij}  -  \alpha^{(\ell)}_{i1}\psi_1(t_{ij})]^2,$$
 subject to $||\psi_1||^2 = 1.$  

 We recast the above criterion into: 
 \begin{align*}
  &\sum_{i=1}^n\frac{1}{n_i}\sum_{j=1}^{n_i} [y_{ij}  -  \alpha^{(\ell)}_{i1}\psi_1(t_{ij})]^2 \\
  =& \sum_{i=1}^n\frac{1}{n_i}\sum_{j=1}^{n_i} [y_{ij}  -  \alpha^{(\ell)}_{i1}\betabold_1^T \bbold(t_{ij})]^2 \\
   =& \sum_{i=1}^n\sum_{j=1}^{n_i} [\frac{1}{\sqrt{n_i}}y_{ij}  - \betabold_1^T  \frac{1}{\sqrt{n_i}} \alpha^{(\ell)}_{i1}\bbold(t_{ij})]^2,
 \end{align*}
 subject to $\betabold_1^T \Gbold \betabold_1 = 1$,
 in which  $\Gbold$ is a $L\times L$ matrix with the $(i,j)$-th element  $\langle b_i,b_j\rangle$. This is a constrained least squares problem.  
Fortunately, we can ignore the norm constrain and obtain the unconstrained least squares minimizer first and then scale it such that its norm is 1.
 % We apply the augmented Lagrange method proposed by \cite{ye1988interior} to numerically solve it. 
More specifically, the solution can be written as 
 $\betabold^{(\ell+1)}_1 = \tilde{\betabold}^{(j+1)}_1/\sqrt{\{\tilde{\betabold}^{(j+1)}_1\}^T\Gbold\tilde{\betabold}^{(j+1)}_1}$, in which $\tilde{\betabold}^{(j+1)}_1 = ({\abold^{(\ell)}}^T\abold^{(\ell)})^{-1}(\abold^{(\ell)})^T\ybold_w $, $\ybold_w = (\ybold_1^T/\sqrt{n_1},\ldots,\ybold_n^T/\sqrt{n_n})^T$ and $ \abold^{(\ell)} = ({\abold_1^{(\ell)}}^T,\ldots,{ \abold_n^{(\ell)}}^T    )^T$ is a $(\sum_{i=1}^n n_i) \times L$ matrix,
 in which $\abold_i^{(\ell)}$ is a $n_i\times L$ matrix with $(p,q)$ elements being  $\alpha^{(\ell)}_{i1}\psi_q(t_{ip})/\sqrt{n_i}.$ It can be checked that the minimizer obtained from the least squares will satisfy the Karush-Kuhn-Tucker condition, thus it is the global minimizer of the loss function \eqref{loss1}

 \item Repeat Step II and III until the algorithm converges. 
 \end{enumerate}

\subsection{Estimating the First and Second FECs\label{k2}}
The first and second FECs are estimated by minimizing
\begin{align*}
\frac{1}{n} \sum_{i=1}^n\frac{1}{n_i}\sum_{j=1}^{n_i} [y_{ij}  -  \alpha_{i1}\psi_1(t_{ij})-\alpha_{i2}\psi_2(t_{ij})]^2\,,
 \end{align*}
subject to $\langle \psi_m,\psi_{\ell} \rangle = \delta_{m \ell}, m, \ell \in \{1,2\}$.  We propose to use the following algorithm to simultaneously estimate $\psi_1(t)$ and $\psi_2(t)$. 

\begin{enumerate}[Step I:]
    \item Set an initial value of $\psi_1^{(0)}(t)$, which can be obtained using the algorithm described in the previous subsection.  
    \item Given the current value of $\psi_1^{(\ell)}(t)$,  we apply the following iterative algorithm to obtain the estimates for $\psi^{(\ell)}_2(t)$ and $\alphabold^{(\ell)}_{m}, m =1,2,$ by minimizing
    $$\frac{1}{n} \sum_{i=1}^n\frac{1}{n_i}\sum_{j=1}^{n_i} [y_{ij}  -  \alpha_{i1}\psi_1^{(\ell)}(t_{ij})-\alpha_{i2}\psi_2(t_{ij})]^2\,,$$
    subject to $\langle \psi_m,\psi_{\ell} \rangle = \delta_{m \ell}, m, \ell \in \{1,2\}$. We apply a similar procedure as described in the previous subsection to obtain the estimates for $\psi^{(\ell)}_2(t)$ and $\alphabold^{(\ell)}_{m}, m =1,2,$ as follows: 
\begin{enumerate}[(1)]
    \item Set an initial value for $\psi_2(t)$, denoted as $\psi^{0}_{2}(t)$, which satisfies $||\psi^{0}_{2}||^2=1$ and $\langle \psi^{0}_{2},\psi^{(\ell)}_1\rangle  = 0;$
    \item Given the current value of $\psi^{(\ell)}_{2}(t)$, we obtain the estimate for $\alphabold^{(\ell)}_{m}, m =1,2$, by minimizing 
    $$\frac{1}{n} \sum_{i=1}^n\frac{1}{n_i}\sum_{j=1}^{n_i} [y_{ij}  -  \alpha_{i1}\psi_1^{(\ell)}(t_{ij})-\alpha_{i2}\psi^{(\ell)}_2(t_{ij})]^2.$$
    This is simply a least squares problem.  For the $i$-th subject, the corresponding $\alphabold^{(\ell)}_i = (\alpha^{(\ell)}_{i1}, \alpha^{(\ell)}_{2i})^T$ is given as 
    $$\alphabold^{(\ell)}_i =(\psibold_{i}^T\psibold_{i})^{-1}\psibold_{i}^T\ybold_i, $$
     where  $\psibold_{i} = (\psibold^{(\ell)}_{i1}, \psibold_{i2}^{(\ell)})$, $\psibold^{(\ell)}_{i1} = (\psi^{(\ell)}_1(t_{i1}),\ldots, \psi^{(\ell)}_1(t_{in_i}))^T$,   $\psibold^{(\ell)}_{i2} = (\psi^{(\ell)}_2(t_{i1}),\ldots, \psi^{(\ell)}_2(t_{in_i}))^T$ and  $\ybold_i = (y_{i1},\ldots, y_{in_i})^T$.

    \item Given the value of $\alphabold^{(\ell)}_{i}$, update the value of $\psi^{(\ell+1)}_{2}(t)$ by minimizing
    $$ \frac{1}{n} \sum_{i=1}^n\frac{1}{n_i}\sum_{j=1}^{n_i} [y_{ij}  -  \alpha^{(\ell)}_{i1}\psi_1^{(\ell)}(t_{ij})-\alpha^{(\ell)}_{2i}\psi_2(t_{ij})]^2$$ 
    subject to $\langle \psi^{(\ell+1)}_{2},\psi_1^{(\ell)} \rangle = 0$ and $||\psi^{(\ell+1)}_{2}||^2 = 1$. Because the norm of $\psi^{(\ell+1)}_{2}(t)$ will not affect the KKT conditions, we can first ignore the norm constraint and the minimization becomes a least square with equality-constraints problem. This problem can also be solved efficiently using the Least Squares with Equalities and Inequalities(LSEI) algorithm proposed by \cite{lawson1995solving}. Then, we normalize the resulting solution such that the norm of $\psi^{(\ell+1)}_{2}(t)$ is 1. 
    \item Repeat step (2) and step (3) until the convergence reaches.      
    \item In the end, we obtain the estimate $\psi^{(\ell)}_2(t)$ and $\alphabold^{(\ell)}_m, m =1,2,$ for the given value of $\psi^{(\ell)}_1(t)$\,.
\end{enumerate}
 \item Given the estimated value $\psi^{(\ell)}_2(t)$, we treat $\psi_1(t)$ as an unknown function and apply the same algorithm within Step II to obtain the estimate for $\psi^{(\ell+1)}_1(t)$ and $\alphabold^{(\ell+1)}_m, m =1,2$. 
 \item Repeat Step II and III until the algorithm converges.

\end{enumerate}

\subsection{Estimating More FECs }
Given the estimates for the first $M$ FECs, $\widehat\psi_i(t), i = 1,\ldots, M$, we can obtain the estimate for $\psi_{M+1}(t)$ and the corresponding ${\alphabold}_{M+1}$ using a similar strategy as described in Subsection \ref{k2}. To be more specific, we iterate between ${\alphabold}_1,{\alphabold}_2,\ldots, {\alphabold}_{M+1}$ and $\psi_{M+1}(t)$ by treating the first $M$ FECs fixed. After we obtain the estimate for $\psi_{M+1}$,  we can further refine those estimates for the first $M$ FECs iteratively by treating each of them as unknown at each iteration.  In this way,  the loss function decreases in the loss function in every iteration. As well, the estimated FECs are always orthogonal to each other.

\subsection{Smoothness Regulation}
In order to control the smoothness of the estimated FECs $\psi_m(t),m=1,\ldots, M$, we can add a roughness penalty in \eqref{sampleloss}. That is, for any fixed $M$, we estimate $\psi_1(t),\ldots,  \psi_M(t)$ by minimizing
\begin{align}
\frac{1}{n} \sum_{i=1}^n\frac{1}{n_i}\sum_{j=1}^{n_i} \bigg[y_{ij} - \sum_{m=1}^M \alpha_{im} \psi_m(t_j)\bigg]^2+\sum_{m=1}^M \gamma_m \int {\bigg[\frac{d^2\psi_m(t)}{dt^2}}\bigg]^2dt, \label{sampleloss2}
\end{align}
subject to $\langle \psi_m,\psi_{\ell} \rangle = \delta_{m \ell}$, where $m, \ell = 1,\ldots, M$. The algorithm introduced in Subsection 3.1-3.3 can be modified accordingly. For instance, we can estimate the first FEC by modifying Step III in Subsection 3.1 as: 
\begin{enumerate}
\item[Step III (b)] Given the current value of $\alphabold^{(\ell)}_1$, we update the estimate of $\psi^{(\ell)}_1(t)$ to  $\psi^{(\ell+1)}_1(t)$ by minimizing 
 $$\frac{1}{n} \sum_{i=1}^n\frac{1}{n_i}\sum_{j=1}^{n_i} [y_{ij}  -  \alpha^{(\ell)}_{i1}\psi_1(t_{ij})]^2 + \gamma_1\int {\bigg[\frac{d^2\psi_1(t)}{dt^2}}\bigg]^2dt,$$
 subject to $||\psi_1||^2 = 1.$  
\end{enumerate}
The above minimization is essentially a quadratically constrained quadratic program(QCQP) problem. We use the R package \texttt{Rsolnp} \citep{Rsolnp} based on the SOLNP algorithm proposed by \cite{Ye1987} to numerically solve it. We will demonstrate the performance of this method in our simulation studies. 

When estimating each FEC, there is only one tuning parameter involved, i.e., the smoothing parameter $\gamma_m$. The value of $\gamma_m$ controls the amount of smoothness imposed on the $m$-th FEC. We propose to select the tuning parameter based on the leave-one-curve-out cross validation strategy. To be more specific, we treat one curve's observations as the test data set and the data for all other curves as the training data set. For instance, when we estimate the first FEC $\psi_1(t)$, we can first obtain the estimate for the first FEC, $\widehat{\psi}\,^{(-i)}_1(t)$, using all the training data for any given value of $\gamma_1$, where we suppose to use the $i$-th curve as the test data set. Then, the score for the test curve can be calculated by minimizing 
$$\sum_{j=1}^{n_i} (y_{ij} - \alpha_{i1}\widehat{\psi}\,^{(-i)}_1(t_{ij}))^2.$$
Then the prediction for $y_{ij}$ is $ \hat{y}_{ij}^{(-i)} = \widehat{\alpha}\,^{(-i)}_{i1}\widehat{\psi}\,^{(-i)}_1(t_{ij})$. The prediction error for the $i$-th curve is 
$$\frac{1}{n_i}\sum_j (\hat{y}\,^{(-i)}_{ij} -  y_{ij})^2. $$
The cross validation error for $\gamma_1$ is given as 
$$\mbox{CV}(\gamma_1) = \sum_{i=1}^n \frac{1}{n_i}\sum_{j=1}^{n_i} (\hat{y}\,^{(-i)}_{ij} -  y_{ij})^2. $$
For the following FEC, we propose to select the smoothing parameter after treating the previous estimated FECs fixed. 
\subsection{Selecting the Number of FECs}
We use the AIC criterion proposed by \cite{li2013selecting} to select the number of FECs: 
$$\mbox{AIC}(M) = N\log(\sigma^2_{M}) + N + 2nM,$$
in which $M$ denotes the number of FECs, $n$ denotes the number of individual curves, and $N=\sum_{i=1}^n n_i$ is the total number of observations. We can estimate the noise variance, $\sigma^2_{M}$, by using the average square of the residuals. That is, 
\begin{align}
\label{AIC}
\hat{\sigma}^2_{M} = \frac{1}{n}\sum_{i=1}\frac{1}{n_i} (\ybold_i - \widehat\ybold_{i,M})^T(\ybold_i - \widehat\ybold_{i,M}), 
\end{align}
{where $\widehat\ybold_{i,M} = (\hat y_{i1},\ldots,\hat y_{in_i})^T$ represents the fitted $i$-th individual's observations when the number of FECs is selected to be $M$. }
\section{Theoretical Results}
Theorem 2 shows that our first estimated FEC will asymptotically converge to the true FEC as the number of subjects increases. Similar results are shown in Theorem 3 for the rest estimated FECs.

{Consider sparse observations of functional data
$ y_{ij} = x_i(t_{ij}) + \epsilon_{ij}$, where
the observation times $ t_{ij}, j = 1, \ldots, n_i, $
for subject $i$
are uniformly drawn from $[0,1]$. Let the Mercer representation for the uncentered covariance function $K(s,t) = \E (X(s)X(t))$ of the stochastic process
$X(t)$ be 
$$K(s,t) = \sum_{m=1}^{\infty} \lambda_m \psi^0_m(s)\psi^0_m(t). $$ Assume $\sum_m \lambda_m < \infty$ and $\int_0^1 [\psi_m^0(t)]^4 d\,t < \infty $
for each $ m = 1,2,\cdots $. 

% \textcolor{red}{KL to mercer}
\begin{thm}
Recall the objective function
\begin{equation}
\label{eq:loss}
L_n(\alphabold, \psi)  =\frac1n \sum_{i=1}^n \frac1{n_i} \sum_{j=1}^{n_i} [ 
y_{ij} - \alpha_{i1} \psi_1(t_{ij})]^2
\end{equation}
where $\alphabold_1=(\alpha_11,\ldots,\alpha_{n1}) \in \mathbb{R}^n $ and $\psi_1(t)$ is a function in $L^2(0,1)$
with constraint $ \int_0^1 \psi^2(t) d\,t = 1$. Then the minimizer $\hat \psi_1(t)$ of $L_n$ converges to $\psi^{0}_1(t)$ in $L^2(0,1)$
almost surely as $n \to \infty$.
\end{thm}

% {\color{red} It will be better to change $ \phi_n(t)$ to $\psi_1(t)$?}

\begin{thm}
The minimizers $\hat\psi_l(t)$, $l=1,\cdots, M$, of the loss function
\begin{equation}
\label{eq:loss2}
L_n(\{\alphabold_l, \psi_l\}_{l=1}^M)  = \frac1n \sum_{i=1}^n \frac1{n_i} \sum_{j=1}^{n_i}
[y_{ij} - \sum_{l=1}^M \alpha_{il} \psi_l(t_{ij})]^2
\end{equation}
converges to
$\psi^0_l(t)$, $l=1,\cdots,M,$ in $L^2[0,1]$ almost surely as $ n \to \infty$.
\end{thm}
}

% Indeed, if we denote $\beta_l$ as the coefficients of projection of $\phi_l(t)$ on the basis $(\psi_k)_k$ of $L^2[0,1]$, then the objective 
% function \eqref{eq:loss2} becomes
% $$L_n(\{\alpha_l, \beta_l\}_{l=1}^M) = \frac1n \sum_{i=1}^n \sum_{k=1}^\infty
% (a_{ik} - (\sum_{l=1}^M\alpha_l \otimes \beta_l)_{ik})^2 m_{ik} + o_P(1)$$ which have
% minimizers $\hat \beta_l$ such that $\hat \beta_l \to e_l $ in $\ell_2$ for each $l\leq M$ by the same argument as for $M = 1$.

The proofs for Theorems 2 and 3 are available in the supplementary documents. The following lemmas are used to prove the above theorem.

\begin{lemma}
\label{lem:slln}
Let $m_i, i=1,2,\ldots,$ be independent positive random variables with mean $1$ and
$ \sum_{i=1}^\infty \E (m_i - 1)^2/i^2 < \infty $.
For any sequence of positive numbers, $a_i$, such that
$ \sum_{i=1}^\infty \E (m_i - 1)^2 a_i^2 / i^2 < \infty$,
we have
\[
\lim_{n \to \infty} \frac1n \sum_{i=1}^n a_i =
\lim_{n \to \infty} \frac{1}{ \sum_{i=1}^n m_i} \sum_{i=1}^n m_i a_i \quad a.s.
\]
\end{lemma}

\begin{lemma}
Let $m_{ij}, i=1,2,\ldots, j=1,2,\ldots, $
be positive random variables.
For each $j=1,2,\ldots$,
$m_{ij}, i=1,2,\ldots$, are independently and 
identically distributed with mean $1$ and finite variance. 
Then for any infinite matrix $A = [A_{ij}]$, 
with $\lambda_j = \lim_n \frac1n\sum_{i=1}^n a_{ij}^2$ exists for each $j$,
as $n \to \infty$,
\[ \lim_{n \to \infty} \frac1n \sum_{i=1}^n\sum_{j=1}^\infty a_{ij}^2 = \lim_{n \to \infty} \frac1n \sum_{i=1}^n\sum_{j=1}^\infty  a_{ij}^2 m_{ij}
=\sum_{j=1}^\infty \lambda_j.
\]
\end{lemma}

\begin{lemma}
For an $n \times p$ matrix $ A$,
the r-rank  approximation 
of $A$ under the Frobenius norm of matrices is
$ \tilde A = \sum_{i=1}^r \alphabold_i \otimes \betabold_i $
where $ r \leq \min(n,p)$, and $\alphabold_i$, $\betabold_i$
are the eigenvectors of $ AA^T$ and $A^T A$, respectively.

% For a matrix $ A_{n \times p} = [ a_{ij} ] $,
% the r-rank with $ r \leq \min(n,p)$  approximation 
% of $A$ under the Frobenius norm of matrices is
% $ \tilde A = \sum_{i=1}^r \alpha_i \otimes \beta_i $
% where $\alpha_i$ and $\beta_i$
% are 
% the r left and r right singular eigenvectors of 
% A; i.e. the eigenvectors of $ AA^T$ and $A^T A$.

\end{lemma}
% }
% \begin{proof}
% By singular value decomposition of $A$, we can
% write $A = \sum_{k=1}^{\min(n,p)} d_k u_k \otimes v_k $, 
% where $(u_k)_{k=1}^n $ and $ (v_k)_{k=1}^p$ are othogonal
% basis in $\mathbb{R}^n$ and $ \mathbb{R}^p $.
% Hence $ A - \sum_{i=1}^r \alpha_i \otimes \beta_i $ 
% has minimum squared Frobenius norm
% $ \sum_{i=r+1}^{\min(n,p) }d_i^2 $ with 
% minimizing $\alpha_i = d_i u_i$ and $ \beta_i = v_i$.
% \end{proof}

%\begin{remark}
%If $A$ has orthogonal columns, as in 
%the score matrix of FPCA,
%$A^TA$ has unit basis vectors as eigenvectors;
%i.e., the $\beta_i$'s are $e_i \in \mathbb{R}^p$, 
%where $e_i$ has only the $i$-th entry non-zero, with value 1.
%\end{remark}

\section{Application: Longitudinal CD4 Percentages}

\begin{figure}[htbp]
    \centering
    \includegraphics[width=\textwidth]{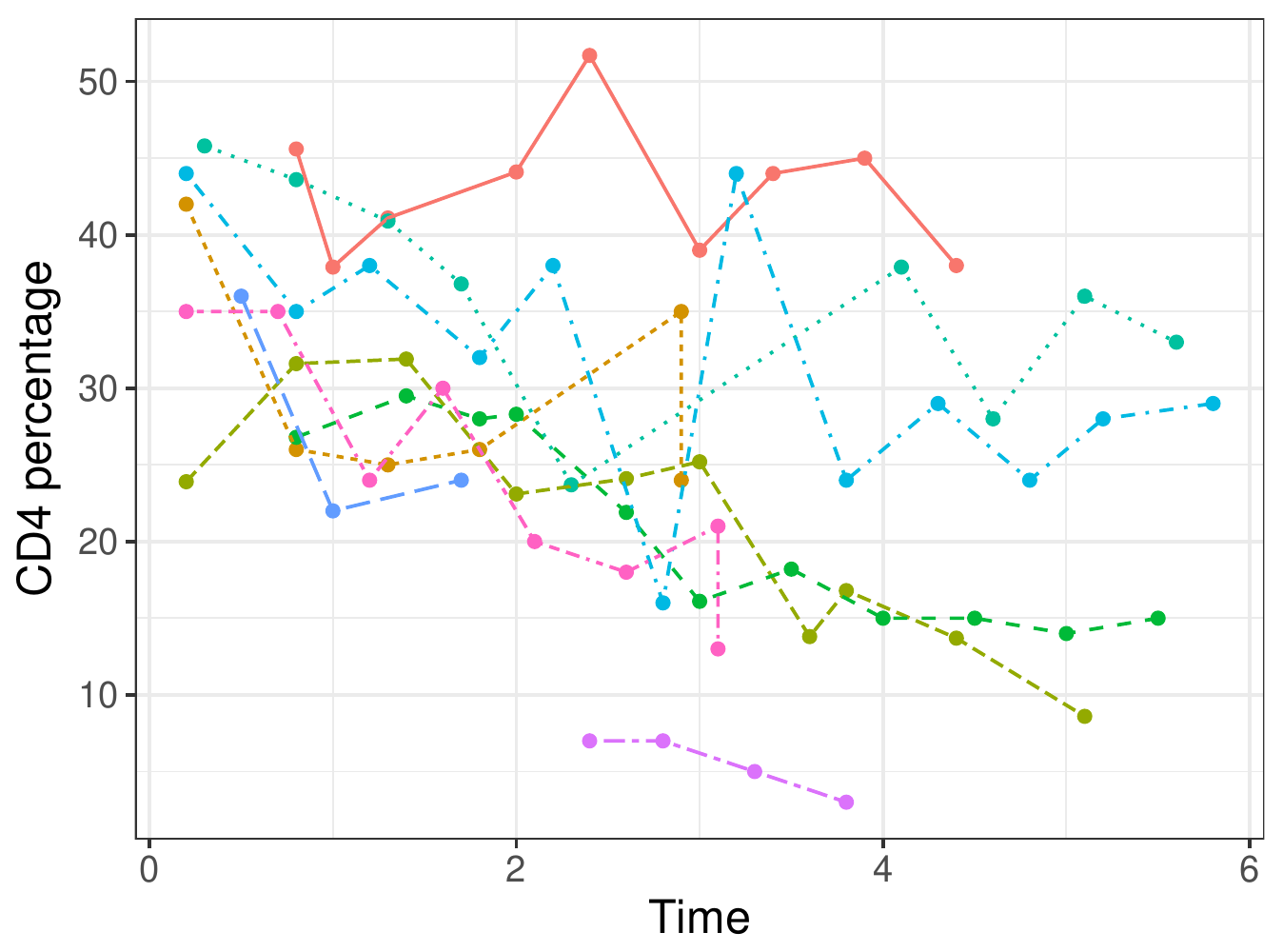}
    \caption{The longitudinal CD4 percentage for 10 randomly selected subjects. Each curve represents the measurements for one single subject.}
    \label{fig:figure1} 
\end{figure}

We demonstrate our proposed method by analyzing the longitudinal CD4 counts dataset. The CD4 percentage, which is defined as CD4 counts divided by the total number of lymphocytes, is a commonly
used marker to describe the health status of HIV infected persons. The dataset considered here is
from the Multi-center AIDS Cohort Study, which includes repeated measurements of CD4 percentages for 283 homosexual men who became HIV positive between 1984 and 1991. All subjects were scheduled to be measured at semi-annual visits. The trajectories of 10 randomly selected subjects are shown in Figure \ref{fig:figure1}. It shows that the data are sparse with unequal numbers of repeated measurements and different visit times for individual subjects, because many of them missed scheduled visits and the HIV infections could occur randomly during the study. For all 283 subjects,  the number of observations per subject ranged between 1 and 14, with a median of 6 measurements. 

\begin{table}[htbp]
\centering
\caption{The values of AIC defined in \eqref{AIC} for various number of FECs.  \label{tab0} }
\begin{tabular}{rrrrrrr}
  \hline
\# FECs & 1 & 2 & 3 & 4 & 5 & 6 \\ 
  \hline
AIC & 8493.44 & 7632.86 & 7626.01 & 7720.19 & 7913.83 & 8059.46 \\ 
   \hline
\end{tabular}
\end{table}

The objective of our analysis is to recover individual longitudinal trajectories from the sparse and irregular observations. The smoothing parameters are selected from $\{0,10^2, 10^4, 10^8\}$ using the leave-one-curve-out cross-validation and the selected smoothing parameters for the first 5 estimated eigenfunctions are $10^4,  10^2, 10^4, 10^2$ and $10^4$, respectively. Table \ref{tab0} displays the values of AIC defined in \eqref{AIC} varying with the number of FECs. It shows that AIC is minimized when the number of FECs is 3. 

\begin{figure}[htbp]
    \centering
    \includegraphics[width=\textwidth]{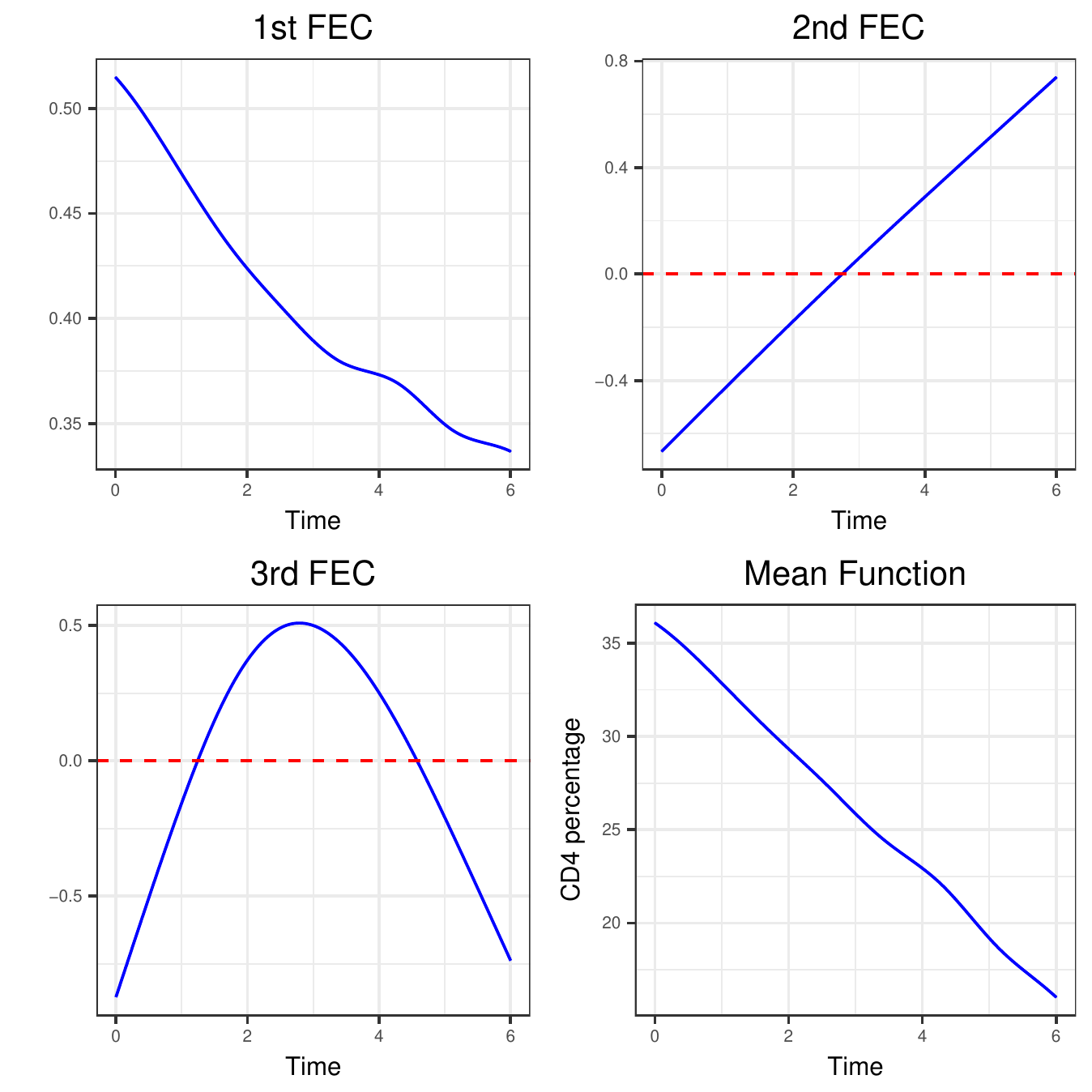}
    \caption{The estimated three functional empirical components (FECs) along with the estimated mean function for the CD4 data.}
    \label{fig:figure2}
\end{figure}

 \begin{figure}[htbp]
    \centering 
    \includegraphics[width=\textwidth]{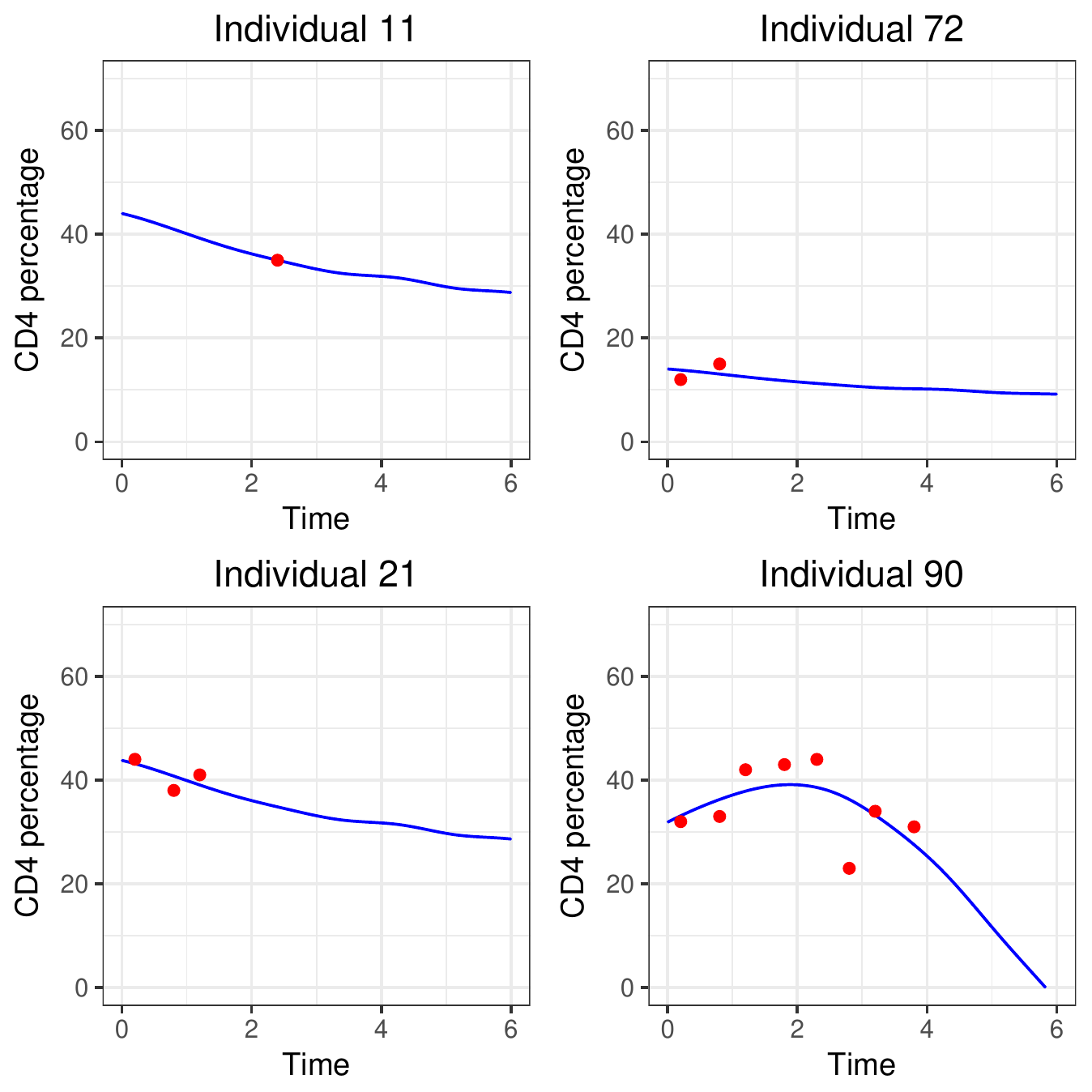}
    \caption{The estimated individual trajectories using the SOAP method (solid line) and the corresponding observations (dots) for individual 11, 21, 72 and 90.}
    \label{fig:figure3}
\end{figure}

 \begin{figure}[htbp]
    \centering   
    \includegraphics[width=0.7\textwidth]{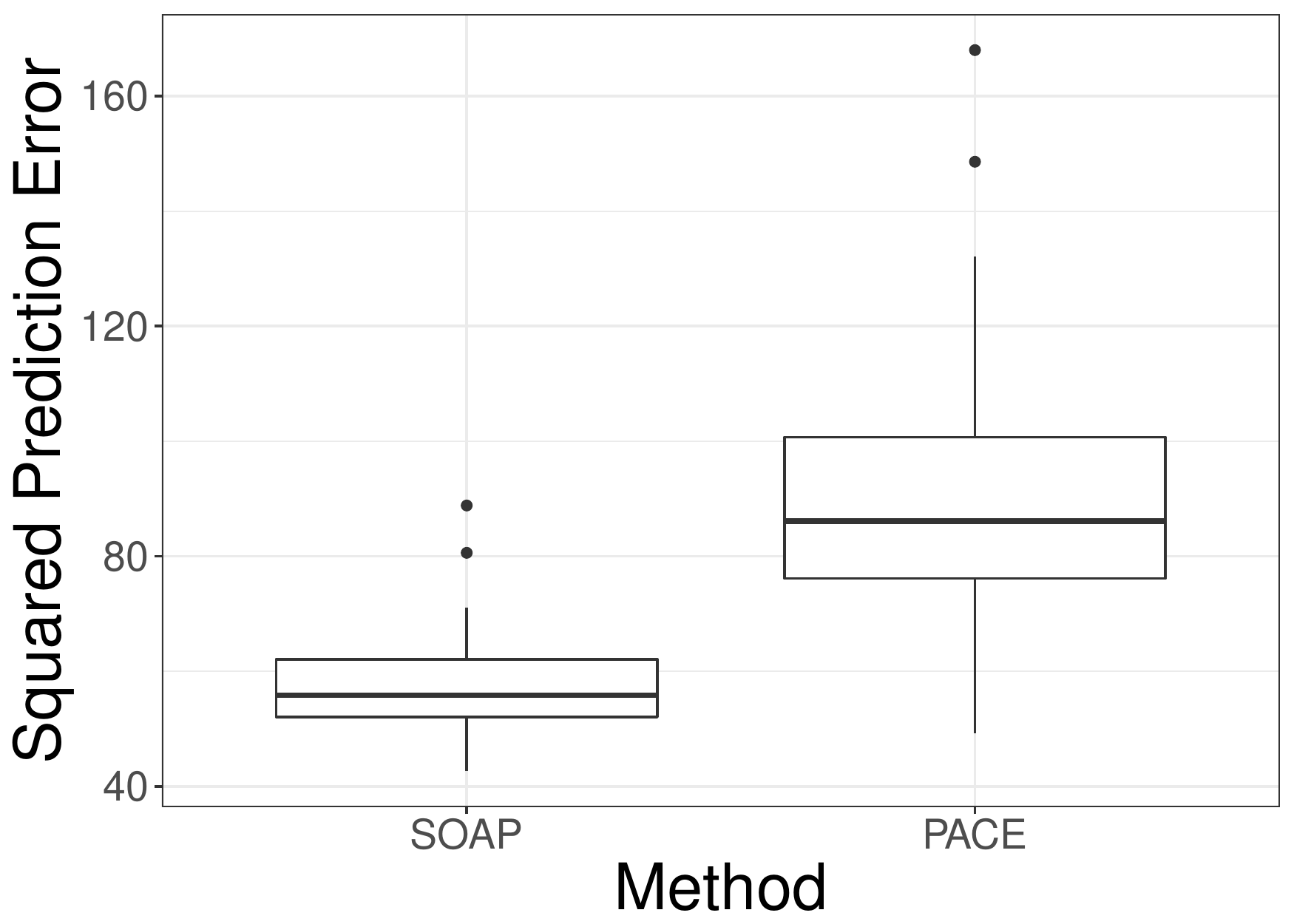}
    \caption{Boxplots of the mean square prediction errors for the last observation in the CD4 dataset using the SOAP method and the PACE method in 100 random data-splitting repetitions.}
    \label{fig:figure4}
\end{figure}

Figure \ref{fig:figure2} shows the estimated three FECs and the estimated mean function. 
The first estimated FEC, $\hat\psi_1(t)$, is decreasing and positive over the whole time interval. The first FEC score can be interpreted as the weighted average of the longitudinal trajectory across time. The second estimated FEC, $\hat\psi_2(t)$, changes its sign at time 3. The second FEC score can be interpreted as the change of the longitudinal trajectory between  $[0, 3]$ and $[3,6]$. Similarly, the third estimated FEC, $\hat\psi_3(t)$, is positive $[1.6,4.3]$ and negative elsewhere. So the third FEC score represents the change of the longitudinal trajectory between $[1.6,4.3]$ and the other periods. The mean function is obtained by taking the average of all the individual predicted trajectories, which shows an overall decreasing trend across individuals. 

 % {\color{red} Can you say some words about Figure  \ref{fig:figure2}? If not, we can move it to the supplementary document? In Figure 2-3, can we add the PACE results? }

Figure \ref{fig:figure3} shows the predicted individual trajectories for 4 different individuals with the various number of observations. It shows that all the estimated CD4 trajectories fit the observations well. An estimated individual trajectory generally displays the overall decreasing trend when the number of observations is small. On the other hand, when there are enough observations for individuals, such as individual 90 shown in Figure \ref{fig:figure3}, the estimated individual trajectory is able to capture the individual trend.

To compare the SOAP method with the PACE method \citep{yao2005functional} with respect to recovering the underlying trajectories, we use the following procedure. First, we randomly select the data of half subjects as the training data set and treat the other half data as the test data set. We estimate the FECs using the training dataset. Next, for each subject in the test data set, we treat the last observation as unknown and predict it based on the previous observations. In the end, we compare the predicted value with the observed value and obtain the mean square prediction error (MSPE) for all individuals in the test data set. We repeat this procedure 100 times.  Figure \ref{fig:figure4} are the boxplots of the MSPEs for two methods. Figure \ref{fig:figure4} shows that the SOAP method outperforms the PACE method in predicting the future individual trajectories. For instance,  the median of MSPEs is 55.87 for the SOAP method, which is 35\% smaller than the PACE method. The 25\% and 75\% quantiles of MSPEs are 52.05 and 62.08 for our method, which are also 31\% and 38\% smaller than the PACE method, respectively.  

% \begin{table}[htbp]
% \centering
% \begin{tabular}{rrrrrrrr}
%   \hline
%  & Mean & Stdev & Median & 1. Quartile & 3. Quartile & Maximum & Minimum \\ 
%   \hline
% SOAP & 57.04 & 7.37 & 55.87 & 52.05 & 62.08 & 88.82 & 42.64 \\ 
%   PACE & 88.65 & 19.19 & 86.12 & 76.15 & 100.68 & 167.99 & 49.19 \\ 
%    \hline
% \end{tabular}
% \end{table}

\section{Simulations}
To evaluate the performance of our proposed method, we conduct one simulation study in comparison with the PACE method. In order to make our proposed method and the PACE method comparable, we simulate the curves $X_i(t)$ such that $E(X_i(t))=0$. Then in this simulation setting, the functional principal components (FPCs) in PACE are equivalent to our proposed functional empirical components (FECs). Therefore, for the rest of this section, we unify both of them as eigenfunctions. 

\begin{figure}[htbp]
    \centering   
    \includegraphics[width=\textwidth]{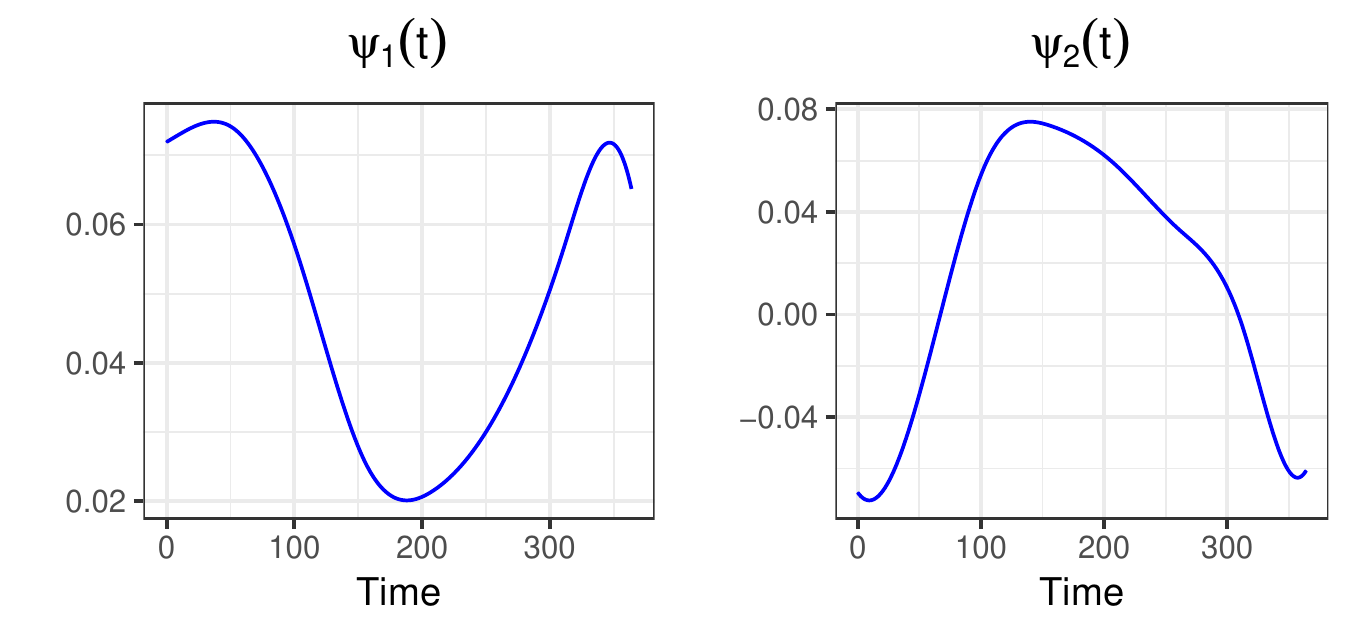}
    \caption{The true eigenfunctions used to generate the true underlying individual trajectories. We obtain these two functional empirical components by conducting conventional FPCA on the Canadian temperature Data \citep{ramsay2002applied}. }
    \label{fig:pcs}
\end{figure}  

The underlying true trajectories are simulated as $X_i(t)  = \alpha_{i1}\psi_1(t)+ \alpha_{i2}\psi_2(t), i=1,\ldots,n$, where the true eigenfunctions, $\psi_1(t)$ and $\psi_2(t)$ are shown in Figure \ref{fig:pcs}, satisfying $\langle \psi_i,\psi_j \rangle =\delta_{ij}, i,j=1,2$.  The corresponding scores $\alpha_{i1}$ and $\alpha_{i2}$ are generated in both Gaussian and non-Gaussian distributions. For the Gaussian scenario, the scores  are generated from two independent Gaussian distributions. That is, $\alpha_{i1}\overset{i.i.d}{\sim} N(0,30)$ and $\alpha_{i2}\overset{i.i.d}{\sim} N(0,10)$.  For the non-Gaussian scenario, the scores are first generated from two independent gamma distributions and then are centered by subtracting the sample mean. That is, $\alpha_{i1}  = \alpha'_{i1} - \bar \alpha'_{i1},$ where $ \alpha'_{i1} \overset{i.i.d}{\sim} \mbox{Gamma}(1,0.03)$ and  $\alpha_{i2}  = \alpha'_{i2} - \bar \alpha'_{i2},$ where $ \alpha'_{i2} \overset{i.i.d}{\sim} \mbox{Gamma}(1,0.1)$. We choose the parameters of these two gamma distributions such that the standard derivations are roughly the same as in the Gaussian scenario. The corresponding observed data for each trajectory are generated as $y_{ij} = X_i(t_{ij}) + \epsilon_{ij}$, in which $\epsilon_{ij}\sim N(0, \sigma^2)$. 
To achieve the sparseness, the number of time points, $n_i$, for each trajectory is chosen randomly from a discrete uniform distribution on $\{1, 2, 3, 4,5\}$ and the corresponding time points $t_{ij}, j = 1,\ldots, n_i$, are uniformly generated in the entire time domain $[0,T]$.

To evaluate the performance of the  SOAP method, we generate 300 training samples and 300 test samples in each simulation replication. We first use our proposed method to estimate the eigenfunctions using the training dataset, and then predict the test samples' trajectories. The PACE method is also applied to estimate the eigenfunctions from the training data and predict the trajectories for the test samples. These two methods are compared by defining the integrated mean prediction error (IMPE) for the 300 test samples as:
\begin{align}\label{mse1}
\mbox{IMPE} = \frac{1}{300}\sum_{i=1}^{300} \int [\widehat{x}_i(t)  -  x_i(t)]^2 dt,
\end{align}
in which $x_i(t)$ represents the true $i$-th trajectory in the test set and $\widehat{x}_i(t)$ is the corresponding predicted trajectory. We repeat the above procedure for 100 repetitions. 

\begin{table}[htbp]
\centering
\caption{The summary results for predicting the individual trajectory using the SOAP method and the PACE method for 100 simulation replicates. The table shows the means, standard derivations (SDs), medians,  minimums and maximums for the integrated mean prediction errors in \eqref{mse1} when the true FPC scores are generated from the Gaussian distribution and non-Gaussian distribution.  \label{tab12} }
\vspace{0.5cm}
\begin{tabular}{rrl|rr}
  \hline
  & \multicolumn{2} {c|} {\bfseries Gaussian}& \multicolumn{2} {c} {\bfseries Non-Gaussian}\\
  \hline
 & SOAP & PACE & SOAP & PACE \\ 
  \hline
  \hline
Mean & 159.38 & $1.02\times 10^5$ & 164.46 & 981.36 \\ 
  SD & 32.45 & $1.18\times 10^6$ & 53.65 & 4.51$\times 10^4$ \\ 
  Median & 154.16 & 151.83 & 151.59 & 290.30 \\ 
  Minimum & 98.42 & 80.87 & 75.81 & 145.45 \\ 
  Maximum & 283.32 & $1.39\times 10^7$ & 521.61 & 5.34$\times 10^5$ \\ 
   \hline
   \hline
\end{tabular}
\end{table}

The results are shown in Table \ref{tab12}. First of all, we find that the performance of PACE is quite unstable when the true FPC scores are generated from the Gaussian distribution and non-Gaussian distribution in comparison with our proposed method. For instance, the maximum $\mbox{IMPE}$ of PACE goes up to $1.3\times 10^{7}$. There are 11 in the Gaussian scenario and 19 in the non-Gaussian scenario out of 100 repetitions that $\mbox{IMPE}$ of PACE is greater than $600$. In contrast, all $\mbox{IMPEs}$ from SOAP are less than $600$.  

\begin{figure}[htbp]
    \centering
    \includegraphics[width=\textwidth]{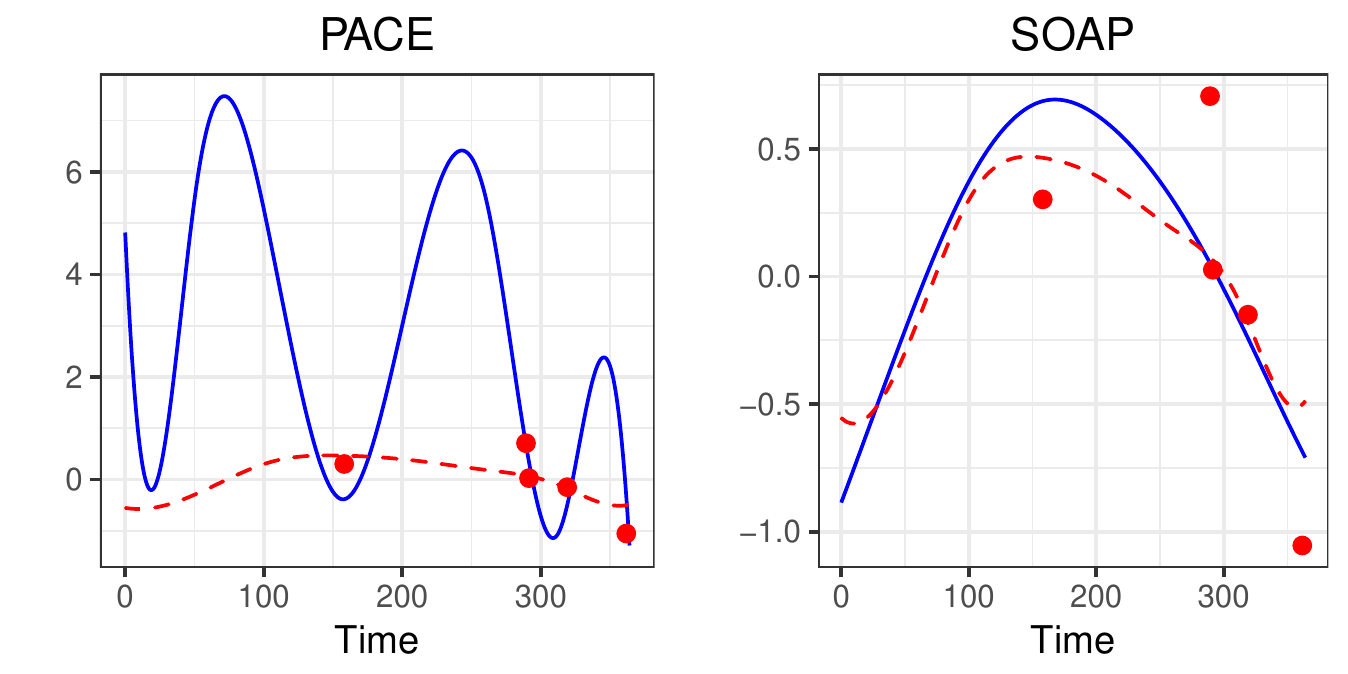}
    \caption{The estimated individual trajectory (solid line) using the PACE method (left panel) and the SOAP method (right panel) compared with the true trajectory (dashed line). The dots represent the observations for this curve. }  
    \label{fig:figure12}
\end{figure}

We notice that PACE produces poorly predicted trajectories when the time points with observations are relatively close to each other. The reason is that estimating the individual scores involves computing the inverse of the sample covariance matrix, and the inverse of this matrix can be quite unstable when the time points with observations are close. Figure \ref{fig:figure12} shows an example from one simulation run with 5 observations in this scenario. Note that those 3 observations in the middle are relatively close in the time domain to each other in comparison with the last observation. As we can see in the left panel in Figure \ref{fig:figure12}, the predicted trajectory produced by PACE overfits the observed data, while the predicted trajectory from SOAP is quite close to the true underlying trajectory.

% \begin{table}[htbp]
% \centering
% \caption{The summary statistics for the prediction errors, i.e., $\mbox{IMSE}_{test}$, in 100 repetitions for the non-Gaussian scenario. SOAP denotes `sparse orthogonal approximation'.  \label{tab1} }
% \begin{tabular}{rrr}
%   \hline
%  & SOAP & PACE \\ 
%   \hline
% Mean & 164.46 & 981.36 \\ 
%   Stdev & 53.65 & 4513.07 \\ 
%   Median & 151.59 & 290.30 \\ 
%   Minimum & 75.81 & 145.45 \\ 
%   Maximum & 521.61 & 53360.79 \\ 
%    \hline
% \end{tabular}
% \end{table}

% \begin{table}[htbp]
% \centering
% \caption{The summary statistics for the prediction errors, i.e., $\mbox{IMSE}_{test}$, in 100 repetitions for the Gaussian scenario. SOAP denotes `sparse orthogonal approximation'.  \label{tab2} }
% \begin{tabular}{rrr}
%   \hline
%  & SOAP & PACE \\ 
%   \hline
% Mean & 159.38 & 101579.62 \\ 
%   Stdev & 32.45 & 1179463.81 \\ 
%   Median & 154.16 & 151.83 \\ 
%   Minimum & 98.42 & 80.87 \\ 
%   Maximum & 283.32 & 13906866.62 \\ 
%    \hline
% \end{tabular}
% \end{table}

\begin{table}[htbp]
\caption{The summary results for estimating the underlying eigenfunctions using the SOAP method and the PACE method for 100 simulation replicates. The table shows the means, standard derivations(SDs), medians,  minimums and maximums for the integrated mean square errors (IMSEs) defined in \eqref{mse2} when the true FPC scores are generated from the Gaussian distribution and non-Gaussian distribution. \label{tab4}  }
\vspace{0.5cm}
\centering
\begin{tabular}{rl|rr|rr}
  \hline
  &&\multicolumn{2}{c|}{\bfseries Gaussian}&\multicolumn{2}{c}{\bfseries Non-Gaussian}\\
  \hline
 & & $\mbox{IMSE}(\hat\psi_1)$ & $\mbox{IMSE}(\hat\psi_2)$ &$\mbox{IMSE}(\hat\psi_1)$ &$\mbox{IMSE}(\hat\psi_2)$ \\ 
  \hline
  \hline
\multirow{2}{*}{\bfseries Mean} & SOAP & 3.20 & 32.05 & 3.56 & 32.90 \\ 
  & PACE & 19.42 & 566.89 & 65.35 & 1012.42 \\ 
  \hline
  \multirow{2}{*}{\bfseries SD}& SOAP & 1.89 & 3.00 & 2.61 & 3.86 \\ 
   & PACE & 12.20 & 451.37 & 56.31 & 551.54 \\ 
   \hline
  \multirow{2}{*}{\bfseries Median} & SOAP & 2.75 & 31.56 & 2.72 & 33.02 \\ 
  & PACE & 15.42 & 406.00 & 46.37 & 961.76 \\ 
  \hline
  \multirow{2}{*}{\bfseries Minimum} & SOAP & 0.47 & 25.81 & 0.45 & 23.59 \\ 
   & PACE & 4.48 & 76.10 & 7.14 & 117.05 \\ 
   \hline
  \multirow{2}{*}{\bfseries Maximum} & SOAP & 9.59 & 40.33 & 14.40 & 44.95 \\ 
   & PACE & 81.98 & 1803.95 & 356.77 & 2157.66 \\ 
   \hline
   \hline
\end{tabular}
\end{table}

Besides recovering the individual trend, we also compare the estimated eigenfunctions with the true eigenfunctions using the following integrated mean square error (IMSE): 
\begin{align}\label{mse2}
\mbox{IMSE}(\hat\psi_i) =\int [\psi_i(t)  - \widehat\psi_i(t)]^2 dt, i =1, 2.
\end{align}
The results are summarized in Table \ref{tab4}. First, the estimated eigenfunctions using the SOAP method are much closer to the true underlying eigenfunctions than those estimated with the PACE method under both simulation settings. For instance, the mean $\mbox{IMSE}(\hat\psi_1)$  from the PACE method is 4 times larger than the SOAP method and $\mbox{IMSE}(\hat\psi_2)$ from the PACE method is almost 18 times larger than the SOAP method. In addition, the performance of the SOAP method is not sensitive to the distribution of the underlying scores,  but the PACE method's performance significantly drops from the Gaussian to the non-Gaussian scenario. For instance, the mean $\mbox{IMSE}(\hat\psi_1)$ increases from 19.42 (Gaussian) to 65.34 (non-Gaussian). Finally, we notice that the performance of the SOAP method is generally more stable than the PACE method, which is shown by comparing the standard derivations of the IMSEs. For example, the standard deviation of $\mbox{IMSE}(\hat\psi_2)$ is 551.54 using the PACE method in comparison with 3.86 using the SOAP method. 

% \begin{table}[htbp]

% \centering
% \begin{tabular}{r|r|rrrrrr}
%   \hline
%  & & & Mean & Stdev & Median & Minimum & Maximum \\ 
%   \hline
%   \multirow{4}{*}{\bfseries Normal}& \multirow{2}{*}{$\mbox{IMSE}(\hat\psi_1)$}& SOAP & 3.20 & 1.89 & 2.75 & 0.47 & 9.59 \\ 
%   & &PACE & 19.42 & 12.20 & 15.42 & 4.48 & 81.98 \\ 
%   \cline{3-8}
%   & \multirow{2}{*}{$\mbox{IMSE}(\hat\psi_2)$} &SOAP & 32.05 & 3.00 & 31.56 & 25.81 & 40.33 \\ 
% & &PACE & 566.89 & 451.37 & 406.00 & 76.10 & 1803.95 \\ 
%    \hline
% \multirow{4}{*}{\bfseries  Non-Gaussian} &    \multirow{2}{*}{$\mbox{IMSE}(\hat\psi_1)$}&SOAP & 3.54 & 2.69 & 2.77 & 0.45 & 14.40 \\ 
%  & & PACE & 65.34 & 57.75 & 46.27 & 7.14 & 356.77 \\ 
%  \cline{3-8}
%  & \multirow{2}{*}{$\mbox{IMSE}(\hat\psi_2)$}& SOAP & 32.90 & 3.86 & 33.02 & 23.59 & 44.95 \\ 
%  & &  PACE & 1012.42 & 551.54 & 961.76 & 117.05 & 2157.66 \\ 
%    \hline
% \end{tabular}
% \end{table}

\section{Conclusions}

In this paper, we propose a novel SOAP method for predicting the underlying individual trajectories as well as the major variation patterns from sparse and irregularly longitudinal observations. The SOAP method directly estimates the empirical functional components from the best approximation perspective. This perspective is different from most conventional methods, such as PACE, which first estimates the de-meaned covariance function from the data and then eigen-decompose the resulting covariance function to obtain the estimated FPCs.  This new best approximation perspective enables the SOAP method to recover the individual trajectories without estimating the mean and covariance functions and without requiring that the underlying FPC scores be Gaussian distributed.   
 
We demonstrate the SOAP method by analyzing a CD4 dataset, in which the longitudinal measurements for each individual are sparsely and irregularly observed. Our SOAP method is able to recover the individual CD4 trajectories and explore the major variational sources across all subjects. We also compare the performance of the SOAP method and the PACE method in prediction by treating the last observation of each individual as unknown and find that the SOAP method produces better predictions compared to the PACE method. 

Furthermore, we evaluate the performance of the SOAP method and the PACE method in a simulation study. We notice that the PACE method can be numerically unstable when the data are observed in close time points. Generally speaking, the SOAP method outperforms the PACE method in both predicting the individual trajectory and recovering the optimal empirical basis functions.

% \begin{table}[htbp]
% \centering
% \caption{The summary statistics for the prediction errors, i.e., $\mbox{IMPE}(\hat\phi_i), i =1,2$, in 100 repetitions for the non-Gaussian scenario. SOAP denotes `sparse orthogonal approximation'.  \label{tab3}  }
% \begin{tabular}{r|rrrrrr}
%   \hline
%  && Mean & Stdev & Median & Minimum & Maximum \\ 
%   \hline
% \multirow{2}{*}{pc1}&SOAP & 3.54 & 2.69 & 2.77 & 0.45 & 14.40 \\ 
%  & PACE & 65.34 & 57.75 & 46.27 & 7.14 & 356.77 \\ 
%  \hline
%  \multirow{2}{*}{pc2}& SOAP & 32.90 & 3.86 & 33.02 & 23.59 & 44.95 \\ 
%  &  PACE & 1012.42 & 551.54 & 961.76 & 117.05 & 2157.66 \\ 
%    \hline
% \end{tabular}
% \end{table}

% \begin{figure}[htbp]
%     \centering
%     \includegraphics[width=\textwidth]{SOAP_sim_EM/SOAP_PACE_pc_fits_Gaussian.pdf}
%     \caption{The estimated FPCs using PACE and SOAP in the Gaussian scenario for 100 simulation repetitions.}
%     \label{fig::figure13}
% \end{figure}
% \begin{figure}[htbp]
%     \centering
%     \includegraphics[width=\textwidth]{SOAP_sim_EM/SOAP_PACE_pc_fits.pdf}
%     \caption{The estimated FPCs using PACE and SOAP in the non-Gaussian scenario for 100 simulation repetitions.}
%     \label{fig::figure14}
% \end{figure}

%\section*{Acknowledgements}
%This research was supported by Nie's Postgraduate Scholarship-Doctorial
%(PGS-D) from the Natural Sciences and Engineering
%Research Council of Canada (NSERC), and the NSERC
%Discovery grant of Cao.

\bibliographystyle{chicago}

 \bibliography{mybib} 

% \bibliography{/Users/joha/Dropbox/Rep_research/latex/mybib} 

\end{document}